\documentclass{article}
\usepackage{amsmath}
\usepackage{chicago}

\newtheorem{theorem}{Theorem}

\newtheorem{corollary}[theorem]{Corollary}

\newtheorem{example}[theorem]{Example}

\newtheorem{lemma}[theorem]{Lemma}

\input{tcilatex}

\begin{document}

\title{Coordination Games With Quantum Information}
\author{Vladislav Kargin \thanks{%
Cornerstone Research, 599 Lexington Avenue floor 43, New York, NY 10022;
skarguine@cornerstone.com}}
\maketitle

\begin{abstract}
The paper discusses coordination games with remote players that have access
to an entangled quantum state. It shows that the entangled state cannot be
used by players for communicating information, but that in certain games it
can be used for improving coordination of actions. A necessary condition is
provided that helps to determine when an entangled quantum state can be
useful for improving coordination.
\end{abstract}

\section{Introduction}

Progress in quantum technology has recently stimulated a surprising
development in game theory. In this development game theory is applied to
conflict situations with the outcome that depends both on participants'
actions and on results of measurements of a quantum state. These conflict
situations has been named quantum games.\footnote{%
See seminal papers by \shortciteN{meyer99} and %
\shortciteN{eisert_wilkens_lewenstein99}, and further development in %
\citeN{johnson01}, \citeN{benjamin_hayden01}, %
\shortciteN{kay_benjamin_johnson01}, \shortciteN{du_li_xu02}, %
\shortciteN{lee_johnson03a}, \shortciteN{lee_johnson03b}, and %
\shortciteN{landsburg04}.}

Quantum games occur in two types: games with quantum strategies and games
with quantum information. In games with quantum strategies actions of
players are operations on a quantum state. A particular emphasis is placed
on quantizations of classical games, which often result in unexpected
outcomes like greater degree of cooperation in quantizations of prisoners'
dilemma.

A different type consists of games with quantum information. These are games
with classical actions, in which players have an access to a shared quantum
state. This quantum state can either encode some useful information or be
used purely for communication and coordination between players. Its
potential usefulness for communication is due to a property of quantum
states that is called entanglement. Entanglement allows measurements on two
remote particles to exhibit a correlated behavior, which cannot be
reproduced using classical correlated random variables.

This paper considers a game with quantum information, in which players need
to coordinate their actions depending on the states of nature that they
privately observe. The players are remote and cannot communicate using
classical channels. They, however, share an entangled quantum state and are
allowed to measure it. The main questions are whether the players can
communicate information about their states of nature and whether they can
use the shared quantum state to coordinate their actions.

It turns out that the players are not able to communicate their private
information but that for certain games they can improve coordination of
their actions. A necessary condition for this to be possible is that the
game should be truly coordination game, that is, that the payoff of the game
should non-trivially depend on both players' states of nature and on both
player's actions.

The rest of the paper is organized as follows: Section 2 provides necessary
background from quantum mechanics and formulates the ``no-signalling''
theorem showing that information cannot be transmitted using an entangled
quantum state. Section 3 introduces a coordination game with quantum
information and defines concepts of entangled and classically generated
signals. Section 4 exhibits an example that shows that entangled signals can
be useful for coordination and describes a necessary condition for their
usefulness. And Section 5 concludes.

\section{Basics of quantum mechanics}

According to quantum mechanics, a quantum state is completely described by a
density matrix. A \emph{density matrix} is a non-negative operator in a
complex Hilbert space that has unit trace. A state with a rank-one density
matrix is called \emph{pure state}. Any state can be represented as a
statistical ensemble of pure states, that is, its density matrix can be
represented as a linear convex combination of rank-one operators.\footnote{%
Good sources of information about modern formulation of quantum mechanics
are \shortciteN{peres95}, \shortciteN{preskill99}, and %
\shortciteN{gill_jupp01}.}

\begin{example}
Qubit
\end{example}

Qubit is a quantum system described by a density matrix in the
two-dimensional Hilbert space. The density matrix can be conveniently
written as a linear combination of the Pauli matrices:%
\begin{equation}
\rho =\frac{1}{2}\left( I+a_{1}\sigma _{1}+a_{2}\sigma _{2}+a_{3}\sigma
_{3}\right) ,
\end{equation}%
where $a_{i}$ are real numbers, and 
\begin{equation}
\sigma _{1}=\left( 
\begin{array}{cc}
0 & 1 \\ 
1 & 0%
\end{array}%
\right) ,\;\sigma _{2}=\left( 
\begin{array}{cc}
0 & i \\ 
-i & 0%
\end{array}%
\right) ,\sigma _{3}=\left( 
\begin{array}{cc}
1 & 0 \\ 
0 & -1%
\end{array}%
\right) .
\end{equation}%
The condition that the density matrix is non-negative restricts the
coefficients: $\sum \left| a_{i}\right| ^{2}\leq 1.$ Therefore the totality
of density matrices corresponds to points inside the unit sphere in
3-dimensional real space. The pure states corresponds to points on the
border of this sphere.

One way to construct a complex system is to join two simpler systems
together. Here is where phenomenon of \emph{entanglement} arises.

Any joint system of two quantum states can be represented by a density
matrix in the product of the Hilbert spaces where each of the states lives.
The simplest possible joint systems are described by Kronecker products of
the density matrices of the parts. We can also take the statistical
ensembles of these joint systems. The density matrix of these more complex
systems is a linear convex combination of the product states. However, some
density matrices in the product space cannot be represented in this way.
These matrices are called \emph{entangled}. The surprising fact is that the
concept of entanglement does not refer to the concept of physical distance
so the parts of an entangled states can be very remote. A measurement on one
part of the state can instantaneously change our knowledge about the state
of the other part. This suggests a question of whether entangled states can
be used for communication.

To address this question, let us first explain how measurements of the
quantum states are described. A \emph{measurement} with a finite number of
outcomes is represented by a collection of non-negative operators $M_{i}$
that adds up to the identity operator. The probability of outcome $i$ is
given by $p_{i}=\mathrm{tr}(M_{i}\rho ).$ In the case with a continuous set
of outcomes $R$ we have a family of non-negative operators $M(x)$ such that 
\begin{equation}
\int_{R}M(x)dx=I.
\end{equation}%
The probability density of outcome $x$ is then given by 
\begin{equation}
p(x)=\mathrm{tr}(M(x)\rho ).
\end{equation}%
This continuous family of operators is often called \emph{Probability
Operator-Valued Measure (POVM)}.

If the system consists of two parts and measurements on the parts are
described by $M(x)$ and $N(y)$, the joint measurement is represented by the
product $M(x)\otimes N(y)$. The outcomes $x$ and $y$ of the joint
measurement are in general correlated. Can we use the correlations for
communication of information? The answer is ``No,'' which can be seen from
the following result.

Suppose that a researcher, Alice, performs one of two possible measurements, 
$M^{(1)}$ or $M^{(2)},$ at her location, and that another researcher, Bob,
performs measurement $N$ at his location. The measurements are performed on
an entangled state described by matrix $\rho $ in the product Hilbert space $%
H_{1}\otimes H_{2}.$ Let for simplicity the measurements have finite sets of
outcomes. Then the joint measurement is represented by product matrices: $%
M_{i}\otimes N_{j}.$

The task of Bob is to determine which measurement, $M^{(1)}$ or $M^{(2)},$
Alice performed. If this were possible, Alice could send information to Bob
by chosing either $M^{(1)}$ or $M^{(2)}.$ However, this is not possible as
the following theorem shows.\footnote{%
This theorem is often called ``no-signalling theorem'' (%
\shortciteN{eberhard78}, \shortciteN{ghirardi_rimini_weber80}, and %
\shortciteN{bussey82}). The version presented here is slightly different in
that it uses the language of operator families to describe measurements.}

\begin{theorem}
\label{no_communication_theorem}Probability of an outcome of measurement $N$
is independent of the choice of measurement $M.$ For any $j:$%
\begin{equation}
\sum_{i}\mathrm{tr}\left( \rho M_{i}\otimes N_{j}\right) =\mathrm{tr}\left[
\rho _{2}N_{j}\right] ,
\end{equation}%
where $\rho _{2}=\left( \mathrm{tr}_{H_{1}}\rho \right) $ is a partial trace
of $\rho $ over $H_{1}.$
\end{theorem}

\textbf{Proof:} Any state is a linear combination of rank-one projectors,
and because of linearity it is enough to prove the theorem for the rank-one
projectors. So assume that\footnote{%
For convenience, we use the Dirac ket-bra notation: the elements of the
Hilbert space are denoted as $\left| e\right\rangle ,$ and the linear
functionals on the Hilbert space are denoted as $\left\langle e\right| .$ In
particular, $\left| e_{0}\right\rangle \left\langle e_{0}\right| $ is the
orthogonal projector on $\left| e_{0}\right\rangle .$ We also write $\left|
e_{i}f_{j}\right\rangle $ instead of $\left| e_{i}\right\rangle \otimes
\left| f_{j}\right\rangle .$} 
\begin{equation}
\rho =\left( \sum a_{ij}\left| e_{i}f_{j}\right\rangle \right) \left( \sum 
\overline{a_{ij}}\left\langle e_{i}f_{j}\right| \right) ,
\end{equation}%
where $\left| e_{i}f_{j}\right\rangle $ is a basis of $H_{1}\otimes H_{2}$.
Then 
\begin{eqnarray}
\sum_{s}\mathrm{tr}\left( \rho M_{s}\otimes N_{t}\right)
&=&\sum_{s}\sum_{ijkl}a_{ij}\overline{a_{kl}}\mathrm{tr}\left( M_{s}\left|
e_{i}\right\rangle \left\langle e_{k}\right| \right) \mathrm{tr}\left(
N_{t}\left| f_{j}\right\rangle \left\langle f_{l}\right| \right)  \notag \\
&=&\sum_{ijkl}a_{ij}\overline{a_{kl}}\left\langle e_{k}\right. \left|
e_{i}\right\rangle \left\langle f_{l}\right| N_{t}\left| f_{j}\right\rangle .
\end{eqnarray}%
This sum clearly does not depend on $M.$ For arbitrary $\rho ,$ it can be
easily evaluated by substituting $M_{s}=I$: 
\begin{equation}
\sum_{i}\mathrm{tr}\left( \rho I\otimes N_{j}\right) =\mathrm{tr}\left[ \rho
_{2}N_{j}\right] .
\end{equation}%
QED.

\section{Coordination Game with Quantum Information}

Suppose player A observe a random state of nature $\varphi $ and player B
observe state of nature $\psi .$ (Here we use the word state in its
classical game-theoretical sense. No confusion should arise with quantum
states.) We assume that states of nature $\varphi $ and $\psi $ are
independent, A cannot observe $\psi $, and B cannot observe $\varphi .$ The
payoffs to players depend both on players' actions, $a$ and $b,$ and on
realization of states. We assume that players maximize the joint payoff that
they can achieve by coordinating their actions, and we will call this game
the \emph{coordination game.}

In case of no communication the strategy of a player can depend only on his
own state. For example, player A's typical strategy is: Play action $a$ with
probability $p_{a}(\varphi ).$ Player B plays action $b$ with probability $%
q_{b}(\psi ).$ A slightly more general situation is when both players
observe a random variable $x,$ which is independents of $\varphi $ and $\psi
.$ Then the players can condition their action on this variable and the
strategies are characterized by probabilities $p_{a}(\varphi ,x)$ and $%
q_{b}(\psi ,x).$

In quantum case the players can perform measurements (which may depend on $%
\varphi $ and $\psi )$ of a shared entangled quantum state and observe
outcomes $s$ and $t.$ Consequently, they can condition their strategies on
these outcomes. The corresponding probability functions are $p_{a}(\varphi
,s)$ and $q_{b}(\psi ,t).$ The essential difference with the classical case
is that $s$ and $t$ are not necessarily independent of $\varphi $ and $\psi
. $

Theorem \ref{no_communication_theorem} shows that entangled states cannot be
used for communication of information. However, this result does not rule
out the possibility that entangled states can be used for coordination
purposes. Thus the question is whether there exist such games, in which
measurements of an entangled quantum state can enhance the joint payoff. A
weaker question is whether there is a couple of random variables $s$ and $t$
that cannot be used for communication but that can increase the payoff in a
coordination game.

More precisely, let us introduce the following definitions. The random
variables $\varphi $, $\psi $, $s,$ and $t$ are \emph{disjoint} if 
\begin{eqnarray}
\Pr \left\{ \psi |\varphi ,s\right\} &=&\Pr \left\{ \psi |\varphi \right\} ,
\label{no_communication_condition} \\
\Pr \left\{ \varphi |\psi ,t\right\} &=&\Pr \left\{ \varphi |\psi \right\} .
\notag
\end{eqnarray}%
In other words, signals $s$ and $t$ do not provide additional information
about $\psi $ and $\varphi $ respectively. The random variables $\varphi $, $%
\psi $, $s,$ and $t$ are \emph{classically generated} if there exists such a
random variable $x$ independent from $\varphi $ and $\psi $ that the
following equality holds for conditional distributions: 
\begin{equation}
p(s,t|x,\varphi ,\psi )=p(s|x,\varphi )p(t|x,\psi ).
\label{def_classical_generation}
\end{equation}%
This means that the pairs of random variables $(s,\varphi )$ and $(t,\psi )$
are independent conditionally on $x.$ For example, the variables are
classically generated if $s$ and $t$ can be represented in the following
form:%
\begin{eqnarray}
s &=&s(\varphi ,x),  \label{classical_generation} \\
t &=&t(\psi ,x)  \notag
\end{eqnarray}%
for some random variable $x$ independent from $\varphi $ and $\psi .$
Classically generated signals are necessarily disjoint.

A quadruple of random variables $\varphi $, $\psi $, $s,$ and $t$ is \emph{%
entangled} if they are disjoint and cannot be classically generated. Abusing
notation we will call signals $s$ and $t$ entangled keeping variables $%
\varphi $ and $\psi $ in the background. We can think about $\varphi $ and $%
\psi $ as the configurations of the measurement apparatuses, and $s$ and $t$
as the outcomes of the measurements. The famous non-locality theorem by Bell
(see for a precise formulation \shortciteN{clauser_horne69}) can be
interpreted as saying that the outcomes of measurements of an entangled
quantum state are entangled in the sense of our definition. However, it is
worth noting that not every quadruple of entangled signals can be realized
by measurements of an entangled quantum state.

\section{Using Entangled Signals for Coordination}

It is surprising but the entangled signals -- although useless for
communication -- can be successfully used for increasing payoff in a
coordination game. Here is a modification of an example due to %
\shortciteN{cleve_watrous04} that shows that measurements of an entangled
quantum state can help in increasing the game payoff.

This example uses only two states per player, which we will take as $\varphi
\in \left\{ 0,\pi /4\right\} $ and $\psi \in \left\{ -\pi /8,\pi /8\right\}
. $ The players have two actions: $0$ and $1$ and their task is to play the
opposite actions unless $\varphi =\pi /4$ and $\psi =-\pi /8,$ in which case
the should play the same action. If they play correctly, then they win and
get 1, otherwise they lose and get zero.

The maximal expected classical payoff is 3/4, which is reached by the
following strategy: Player 1 always plays 0, player 2 always plays 1. \ What
is the optimal quantum strategy?

Suppose that the players share an entangled state with the density matrix,
which is the projector on the following vector:%
\begin{equation*}
\eta =\frac{1}{\sqrt{2}}\left( \left| 0\right\rangle \otimes \left|
1\right\rangle -\left| 1\right\rangle \otimes \left| 0\right\rangle \right) ,
\end{equation*}%
where $\left| 0\right\rangle $ and $\left| 1\right\rangle $ are basis
vectors in a two-dimensional Hilbert space. First, let us calculate the
probabilities of outcomes if the first player measures the state by
projecting it on two orthogonal vectors 
\begin{eqnarray*}
m_{0}(\theta _{1}) &=&\cos \theta _{1}\left| 0\right\rangle +\sin \theta
_{1}\left| 1\right\rangle , \\
m_{1}(\theta _{1}) &=&-\sin \theta _{1}\left| 0\right\rangle +\cos \theta
_{1}\left| 1\right\rangle ,
\end{eqnarray*}%
and the second player measures the state by projecting it on 
\begin{eqnarray*}
n_{0}(\theta _{2}) &=&\cos \theta _{2}\left| 0\right\rangle +\sin \theta
_{2}\left| 1\right\rangle , \\
n_{1}(\theta _{2}) &=&-\sin \theta _{2}\left| 0\right\rangle +\cos \theta
_{2}\left| 1\right\rangle .
\end{eqnarray*}%
(It is easy to see that projectors on a complete system of orthogonal
vectors form a collection of measurement operators as was defined above.)

For example, the probability of outcome 00 is 
\begin{eqnarray*}
p_{00} &=&\mathrm{tr}\left\{ \left( \left| m_{0}\right\rangle \left\langle
m_{0}\right| \otimes \left| n_{0}\right\rangle \left\langle n_{0}\right|
\right) \left| \eta \right\rangle \left\langle \eta \right| \right\} \\
&=&\frac{1}{2}\left( \cos \theta _{1}\sin \theta _{2}-\sin \theta _{1}\cos
\theta _{2}\right) ^{2} \\
&=&\frac{1}{2}\sin ^{2}\left( \theta _{2}-\theta _{1}\right) .
\end{eqnarray*}%
Computing similarly all other probabilities we have the following table:%
\begin{equation*}
\left( 
\begin{array}{cc}
p_{00} & p_{01} \\ 
p_{10} & p_{11}%
\end{array}%
\right) =\frac{1}{2}\left( 
\begin{array}{cc}
\sin ^{2}\left( \theta _{2}-\theta _{1}\right) & \cos ^{2}\left( \theta
_{2}-\theta _{1}\right) \\ 
\cos ^{2}\left( \theta _{2}-\theta _{1}\right) & \sin ^{2}\left( \theta
_{2}-\theta _{1}\right)%
\end{array}%
\right) .
\end{equation*}

Now, suppose that the first and second players use measurements with
parameters $\theta _{1}=\varphi $ and $\theta _{2}=\psi $ respectively. Let
the players play the outcome they observed. Then they play the opposite
actions with probability $\cos ^{2}(\pi /8)$ unless $\varphi =\pi /4$ and $%
\psi =-\pi /8$, $\ $\ in which case they will play it with probability $\cos
^{2}(3\pi /8)=\sin ^{2}(\pi /8).$ Therefore the probability to win is $%
1/4\left\{ 3\cos ^{2}(\pi /8)+(1-\sin ^{2}(\pi /8))\right\} =\cos ^{2}\left(
\pi /8\right) >3/4.$

It turns out that this is the maximal probability of win in this game
achievable by quantum strategies.

Does sharing a quantum state always increase the maximal expected payoff in
coordination games? The answer is ``No''. In some games sharing a quantum
state does not help. Then what are conditions that make the sharing helpful?
One necessary condition is that the payoff must depend on the states of
nature of both players.

We need some preliminary definitions to formulate the theorem. First, let us
for simplicity identify the actions of players with their signals: The
players simply play the signal that they obtained. It can be shown that
every coordination game can be cast in this form. Let the distribution of
states of nature $\varphi $ and $\psi $ be $p(\varphi )p(\psi ),$ where $%
p(\varphi )$ and $p(\psi )$ are marginal distributions of $\varphi $ and $%
\psi .$\footnote{%
Here $p$ simply stands for probability distribution. In particular it may
denote two different distribution functions for $\varphi $ and $\psi .$ A
more precise but cumbersome notation would be $p_{\varphi }(x)$ and $p_{\psi
}(y).$} Let us also use notation 
\begin{equation}
p(s,t,\varphi )=\sum_{\psi }p(s,t,\varphi ,\psi ).
\end{equation}%
We will call signals $s$ and $t$ state-consistent if the $\varphi $-$\psi $
marginal of the distribution $p(s,t,\varphi ,\psi )$ coincide with $%
p(\varphi ,\psi ).$ Then the following theorem holds:

\begin{theorem}
Assume that payoff in a game depends only the first player's state of
nature: $\pi =\pi (s,t;\varphi ).$ Then the maximal expected payoff when
players use state-consistent entangled signals coincides with the maximal
expected payoff when players use certain state-consistent
classically-generated signals.
\end{theorem}

\textbf{Proof:} Let the distribution of state-consistent entangled signals
that maximize payoff be $p(s,t,\varphi ,\psi ).$ Consider the following
distribution:%
\begin{equation}
\widetilde{p}(s,t,\varphi ,\psi )=p(s,t,\varphi )p(\psi ),
\end{equation}%
which is a product of marginal distributions for $(s,t,\varphi )$ and $\psi $%
. It is clear that this new distribution is state consistent: The marginal
distribution $\widetilde{p}(\varphi ,\psi )$ is the same as $p(\varphi ,\psi
).$ Moreover, the marginal distribution $\widetilde{p}(s,t,\varphi )$ is the
same as $p(s,t,\varphi ),$ which implies that the expected payoff based on
the strategy that uses signals distributed according to $\widetilde{p}%
(s,t,\varphi ,\psi )$ is the same as that of strategy that uses signals
distributed according to $p(s,t,\varphi ,\psi ).$ Indeed,%
\begin{eqnarray}
\sum_{\varphi ,\psi }\pi (s,t;\varphi )p(s,t,\varphi ,\psi )
&=&\sum_{\varphi }\pi (s,t;\varphi )p(s,t,\varphi ) \\
&=&\sum_{\varphi ,\psi }\pi (s,t;\varphi )\widetilde{p}(s,t,\varphi ,\psi ).
\end{eqnarray}%
We will prove the theorem by showing that the signals with distribution $%
\widetilde{p}(s,t,\varphi ,\psi )$ can be classically generated.

We need a lemma that claims that conditional distribution of the second
player's signal does not depend on the first player's state.

\begin{lemma}
For entangled signals, the following equality holds for any $\varphi _{1}$
and $\varphi _{2}$: $p(t|\varphi _{1})=p(t|\varphi _{2}).$
\end{lemma}

Proof: By definition of entangled signals (\ref{no_communication_condition})
we have $p(\varphi |t,\psi )=p(\varphi |\psi ).$ Consequently $p(\varphi
|t)=p(\varphi ),$ and%
\begin{equation}
p(t|\varphi )=\frac{p(t,\varphi )}{p(\varphi )}=\frac{p(\varphi |t)}{%
p(\varphi )}p(t)=p(t).
\end{equation}%
The last term in this expression does not \ depend on $\varphi .$ QED.

\begin{corollary}
If $(s,t,\varphi ,\psi )$ is distributed according to $\widetilde{p}%
(s,t,\varphi ,\psi )$ then $t$ is independent of $\varphi $ and $\psi .$
\end{corollary}

\textbf{Proof:} 
\begin{eqnarray}
\widetilde{p}(s,t,\varphi ,\psi ) &=&p(s,t,\varphi )p(\psi ) \\
&=&p(s|t,\varphi )p(t|\varphi )p(\varphi )p(\psi ) \\
&=&p(s|t,\varphi )p(t)p(\varphi )p(\psi ).
\end{eqnarray}%
Summing over $s$ we get 
\begin{equation}
\widetilde{p}(t,\varphi ,\psi )=p(t)p(\varphi ,\psi ).
\end{equation}%
QED.

Because of the Corollary, we can take $t$ as $x$ in the definition of the
classically generated variables (\ref{def_classical_generation}) and we then
get:%
\begin{equation}
\widetilde{p}(s,t|t,\varphi ,\psi )=p(s|t,\varphi )p(t|t,\psi )
\end{equation}%
as required in this definition. QED.

\section{Conclusion}

We discussed the value of quantum entanglement for coordination games with
remote players. We showed that quantum entanglement cannot be used for
communicating information, but that it can be useful for coordination
purposes. We also introduced the concept of entangled signals which is a
weaker concept than the concept of quantum entanglement but captures some of
its properties. This new concept may be helpful in analyzing the properties
of quantum entanglement.

\bibliographystyle{CHICAGO}
\bibliography{comtest}

\end{document}